\documentclass[preprint,secnumarabic,nobibnotes,aps,12pt]{revtex4-1}
\usepackage{amsmath}
\usepackage{amsfonts}
\usepackage{amssymb}
\usepackage[usenames,x11names]{xcolor}
\usepackage{slashed}
\usepackage{graphicx}
\usepackage{array}
\usepackage[pdftex]{hyperref}			% For creating hyperlinks in cross
	\hypersetup{				% references
		colorlinks = true,
		linkcolor = red,
                linktocpage = true,
		citecolor = Blue4,
		urlcolor = Green4
	}

\begin{document}

\title{130 GeV Gamma Ray Signal in NMSSM by Internal Bremsstrahlung}
\author{Gaurav Tomar}
\email{tomar@prl.res.in}
\affiliation{Physical Research Laboratory, Ahmedabad 380009,
India}
\author{Subhendra Mohanty}
\email{mohanty@prl.res.in}
\affiliation{Physical Research Laboratory, Ahmedabad 380009,
India}
\author{Soumya Rao}
\email{soumya@prl.res.in}
\affiliation{Physical Research Laboratory, Ahmedabad 380009,
India}
%%%%%%%%%%%%%%%%%%%%%%%%%%%%%%%%%%%%%%%%%%%%%%%%%%%%%%%%%%%%%%%%%%%%%%%%%%%%%%%%%%%%%%%%%%%%%%%%%%%%%%%%%%%%%%%%%%%%%%%%%%%%%%%
%%%%%%%%%%%%%%%%%%%%%%%%%%%%%%%%%%%%%%%%%%%%%%%%%%%%%%%%%%%%%%%%%%%%%%%%%%%%%%%%%%%%%%%%%%%%%%%%%%%%%%%%%%%%%%%%%%%%%%%%%%%%%%%
\begin{abstract}
There is a possible $\gamma$-ray signal at 130 GeV coming from the Galactic Center as seen by Fermi-LAT experiment. 
We give a SUSY dark matter model to explain this $\gamma$-ray feature in NMSSM. 
We show that in NMSSM, one can have a benchmark set in which the $\gamma$-ray signal arises from final state $\gamma$'s
in the 
$\chi \chi \rightarrow f \bar f \gamma$ annihilation of a 130 GeV bino dark matter requiring a boost factor of 
$\sim 590$ to fit the $\gamma$-ray signal. In addition, this benchmark set also gives the correct relic density, lightest 
Higgs mass of 125 GeV and is consistent with constraints on SUSY from LHC. 
This dark matter model evades the XENON100 constraint but is
testable in a future XENON1T experiment.  
\end{abstract}
\maketitle 
%%%%%%%%%%%%%%%%%%%%%%%%%%%%%%%%%%%%%%%%%%%%%%%%%%%%%%%%%%%%%%%%%%%%%%%%%%%%%%%%%%%%%%%%%%%%%%%%%%%%%%%%%%%%%%%%%%%%%%%%%%%%%%
\section{Introduction}
The minimal supersymmetric standard model (MSSM)\cite{Djouadi2008a} is an extension  of the standard model, in which 
the Higgs mass is protected from receiving large radiative corrections and in addition it has a plethora of weakly interacting
neutral particles, which can serve as candidates for dark matter. MSSM has the problem of explaining the scale of the $\mu$ 
parameter, which
is addressed by extending MSSM with an additional gauge singlet superfield in the so called NMSSM (Next-to-Minimal 
Supersymmetric Standard Model) \cite{DREES1989,Ellwanger2010,Ross2012,Hall2012,King2012,Kang2012,Cao2012,Vasquez2012}, where
the $\mu$ parameter can arise from the vev of the singlet field.\\
The Higgs sector of NMSSM contains three CP-even mass eigenstates, the lowest of which is interpreted as the 125 GeV 
Higgs, discovered at LHC \cite{Aad2012,Chatrchyan2012b}. In NMSSM, Higgs mass gets an additional contribution from the 
singlet-Higgs interaction in the scalar potential, which helps
% and Higgs scalar coupling $\lambda$ has taken to be large 
% \cite{King2012} in order 
to achieve 125 GeV Higgs mass after loop corrections.\\ 
The neutral sector of NMSSM has five mass eigenstates, four gauginos from MSSM and in addition a singlino component.
In NMSSM, there are two extra candidates for dark matter compared to MSSM, namely the scalar singlet and its superpartner 
singlino. A viable dark matter must satisfy the relic density $\Omega_{\chi}h^2 = 0.1199 \pm 0.0027$, 
observed by WMAP-PLANCK \cite{Ade2013,Hinshaw2012} and it should also satisfy the bounds from direct detection experiments like 
XENON100, CDMS \cite{Aprile2012,Ahmed2010,Ahmed2011} etc. There are hints of possible signals of dark 
matter in cosmic ray and like
the 130 GeV gamma ray signal from the Galactic Center, observed by Fermi-LAT experiment \cite{Atwood2009}, 
which can be explained by the dark matter of mass $\sim 130$
GeV and annihilation cross-section $\langle\sigma v\rangle_{\gamma \gamma}=1.27\pm 10^{-27} \rm cm^{3} sec^{-1}$
\cite{Weniger2012,Tempel2012a,Hektor2013a,Su2012,Su2012a,Hooper2012,Hektor2012,Mirabal2013,Zechlin2012}.
There are also present some non dark matter explanations of the 130 GeV gamma ray signal, like instrumental noise 
\cite{Whiteson2012}, statistical fluctuation, Earth-limb magnification \cite{Hektor2012a} and emission from 
AGNs \cite{Mirabal2013}, but in our work, we focused on the dark matter explanation of this signal.
\\  
% In this paper we found a set of bench mark point where the singlet-Higgs coupling term 
% $\lambda \sim -0.62$ can help in achieving 125 GeV Higgs mass without excessive fine tunning and without taking coupling to
% the non perturbative regime.\\
% Recent studies of the Fermi-LAT \cite{Atwood2009} data has reveled an interesting feature in the form of monochromatic line 
% \cite{Tempel2012a,Weniger2012} at 130 GeV coming from the vicinity of galactic center. There are various models 
% \cite{Finkbeiner2013,Hektor2012a,Mirabal2013,Whiteson2012}, which explained this feature as a instrumental, statistical, 
% Earth Limb or AGN effects. 
% This monochromatic line can be interpreted 
% as dark matter annihilation into two photons \cite{Weniger2012} with annihilation cross-section 
% $\langle\sigma v\rangle_{\gamma \gamma}=1.27\pm 10^{-27} \rm cm^{3} sec^{-1}$
% \cite{Tempel2012a,Hektor2013a,Su2012,Su2012a,Hooper2012,Hektor2012,Mirabal2013,Zechlin2012}.
To satisfy Fermi-LAT data \cite{Atwood2009}, we require large dark matter annihilation cross-section with 
$m_{\chi} \sim 130$ GeV \cite{Weniger2012}. Within MSSM, if one 
requires a large annihilation cross-section into photons, it could mean that neutralino dark matter is a higgsino, wino or 
mixture of 
them. However if dark 
matter is a $130 ~\rm GeV$ higgsino or wino, then the annihilation cross-section at freeze out will be large and that would lead 
to a  negligible thermal relic density. In addition a wino or 
higgsino annihilation would give a continuum gamma ray spectrum \cite{Cohen2012}, which constraints these models. 
In \cite{Shakya2012}, a bino
dark matter of 130 GeV produces gamma rays through the internal bremsstrahlung ($\chi \chi \rightarrow f \bar f \gamma$),  
which makes use of well
known fact \cite{Bergstrom1989,Bergstrom2005} that if we add a photon in $\chi \chi \rightarrow f \bar f$ annihilation then 
we can get rid from helicity suppression factor $(m_f/m_{\chi})^2$, but of course paying the price of $\alpha_{em}$ 
suppression.
As it has been pointed out \cite{Shakya2012}, if we 
want a lightest supersymmetric particle (LSP) in the form of pure bino then we would need to make the higgsinos fraction small
by taking a large value of 
$\mu \gtrsim 500 ~\rm GeV$. This leads to the naturalness issue in MSSM, where the $\mu$ parameter is expected to be 
$100-200 ~\rm GeV$. 
A way out is in NMSSM, where an extra contribution comes to the higgsinos masses from singlino-higgsinos mixing. 
Making this mixing
large ensures that higgsinos are heavy and their contribution in LSP become small, keeping this model natural. 
In NMSSM, one can also raise dark matter annihilation cross-section through a resonance channel, for which the mass of pseudo 
scalar tuned as $m_A \approx 2m_{\chi^0}$ \cite{Das2012,Chalons2013}. 
This involves tunning in parameters of two unrelated sector of the model, 
which does not seem natural. So, internal bremsstrahlung in NMSSM provides a more natural way to explain the observed feature of
Fermi-LAT.\\
% In this paper, we study a dark matter model in NMSSM, which would explain the 125 GeV Higgs mass without excessive fine 
% tunning, gives the correct relic density of dark matter, evades the XENON100 \cite{Aprile2012} bound (but testable in 
% XENON1T \cite{Aprile2012a}) and satisfy the constraints from flavour physics 
% \cite{Nakao2004,Barberio2008,Aaltonen2008,Aubert2008a} and LHC. 
% Our model explains the observed 130 GeV line in Fermi-LAT data from the photon emission in dark matter annihilation by internal 
% bremsstrahlung \cite{Bergstrom2008a,Bringmann2012,Bell2011}
In this paper, we studied NMSSM and present a set of bench mark point, in which LSP is a bino with mass $\sim 130$ GeV.  
The correct relic density has been obtained by taking low stau mass and it is achieved by stau-coannihilation. 
In the Higgs sector, one
can have 125 GeV higgs by taking $\hat H_u \cdot \hat H_d \hat S$ coupling $\lambda \sim -0.62$, the same large lambda in 
the neutralino sector 
ensures that the 
higgsino are heavy due to singlino-higgsino mixing and the $\mu$ 
parameter need not to be tuned for the large value of higgsinos components. The gamma ray signal comes by internal 
bremsstrahlung 
$(\chi \chi \rightarrow \bar f f \gamma)$ and therefore we need not tune the 
pseudo scalar mass to twice the bino mass as in the other NMSSM models \cite{Das2012,Chalons2013}, which explain the 130 GeV 
signal.\\
We used micrOMEGAs code 3.1 \cite{Belanger2013a} and computed the gamma-ray flux, electron-positron flux and antiproton 
flux from the annihilation of 130 GeV bino dark matter. We used these fluxes in GALPROP code 
\cite{Strong2007,Moskalenko1998}, following
isothermal dark matter density profile \cite{Bahcall1980} and
found that assuming a boost factor \cite{Bergstrom2008a,Mohanty2012} 
of $\sim 590$, we can explain the 130 GeV gamma-ray signal, while still not exceeding the electron, positron and antiproton 
data \cite{Abdo2009,Ackermann2010,Adriani2009b,Adriani2010}. We also find that our bench mark values of parameters ensure small 
direct detection cross-section, such that there 
is no signal for this dark matter in XENON100 experiment \cite{Aprile2012} but we predict a scattering cross-section of 
O $\sim 10^{-46} \rm cm^2$
, which may be seen in XENON1T experiment \cite{Aprile2012a}.\\
The arrangement of the paper is as follows. In section II, we discussed the general NMSSM model and compared the output of our
model with their Standard Model counterparts. In section III, we discussed
internal bremsstrahlung in detail. After describing the perspective of our bench mark point for the direct and indirect 
detection of dark matter in section IV and V, we conclude in section VI. 
%%%%%%%%%%%%%%%%%%%%%%%%%%%%%%%%%%%%%%%%%%%%%%%%%%%%%%%%%%%%%%%%%%%%%%%%%%%%%%%%%%%%%%%%%%%%%%%%%%%%%%%%%%%%%%%%%%%%%%%%%%%%%%
\section{NMSSM}
NMSSM is a singlet extension of MSSM \cite{Djouadi2008a}, which can solve the $\mu$ problem of MSSM. In this singlet 
extension of MSSM the superpotential can be written as \cite{Ellwanger2010},
\begin{align}
 {\cal W} = \lambda \hat S \hat H_u \cdot \hat H_d + \frac{\kappa}{3} \hat S^3 + h_u \hat Q\cdot \hat H_u \hat U^c_{R}
    +h_d \hat H_d\cdot \hat Q \hat D^c_{R}+h_e \hat H_d\cdot \hat L \hat E^c_{R}
 \label{spo}
\end{align}
where $\lambda$, $\kappa$, $h_u$, $h_d$ and $h_e$ are the dimensionless Yukawa couplings.
In eq.(\ref{spo}), SU(2) doublets are
\begin{eqnarray*}
\hat Q = \left(\begin{array}{cc} \hat U_L \\ \hat D_L\end{array}\right), 
\hat L = \left(\begin{array}{cc} \hat \nu_L \\ \hat E_L\end{array}\right), 
\hat H_u = \left(\begin{array}{cc} \hat H^{+}_u \\ \hat H^{0}_u\end{array}\right), 
\hat H_d = \left(\begin{array}{cc} \hat H^{0}_d \\ \hat H^{-}_d\end{array}\right)
\end{eqnarray*}
soft SUSY breaking terms, which correspond to the masses and couplings of fields, mentioned in eq.(\ref{spo}), are
\begin{align}
 -{\cal L}_{soft} &= m^2_{H_u} |H_u|^2 + m^2_{H_d} |H_d|^2 + m^2_S |S|^2\\ \nonumber
           &+ m^2_{Q} |Q^2| + m^2_{U}|U^2_R| + m^2_{D} |D^2_R|
            + m^2_{L}|L^2| + m^2_{E}|E^2_R|\\ \nonumber
           &+(\lambda A_{\lambda} H_u \cdot H_d S + \frac{1}{3} \kappa A_{\kappa} S^3 + B_{\mu} H_u \cdot H_d \\ 
           &+ h_u A_u Q \cdot H_u { U}^c_R - h_d A_d {Q} \cdot H_d {D}^c_R 
            -h_e A_e L \cdot H_d {E}^c_R + h.c.) \label{sof}
\end{align}
Higgs potential can be obtained from the F, D and soft SUSY breaking terms and it is given as, 
\begin{align}
 V &= |\lambda (H^{+}_u H^{-}_d - H^{0}_u H^{0}_d) + \kappa S^2|^2\\ \nonumber
   &+(m^2_{H_u} + |\mu + \lambda S|^2) (|H^0_{u}|^2 + |H^{+}_u|^2)  
    + (m^2_{H_d} + |\mu + \lambda S|^2) (|H^0_{d}|^2 + |H^{-}_d|^2)\\ \nonumber
   &+ \frac{g^2_{1}+g^2_{2}}{8} (|H^0_u|^2 + |H^+_u|^2 - |H^0_d|^2 - |H^-_d|^2)^2 
    + \frac{g^2_2}{2} |H^+_u H^{0\ast}_d + H^0_u H^{-\ast}_d|^2 \\ \nonumber
   &+ m^2_S |S|^2 + (\lambda A_\lambda(H^+_u H^-_d-H^0_u H^0_d) S + \frac{1}{3} \kappa A_\kappa S^3 
    + B\mu (H^+_u H^-_d-H^0_u H^0_d) \\
   &+ \rm h.c) 
\end{align}
Here $g_1$ and $g_2$ represent $U(1)_Y$ and $SU(2)$ couplings respectively. After expanding Higgs potential
around the real natural vevs $~v_u, ~v_d~\rm and ~s$, considering minimization conditions, Higgs mass matrices can be found, 
see \cite{Ellwanger2010} for more discussion.\\ 
The neutralino mass matrix of NMSSM in the basis 
$(\tilde B,\tilde W^0, \tilde H^0_d, \tilde H^0_u, \tilde S)$ is given by,
\begin{eqnarray}
{\cal M_\chi} = \left(\begin{array}{ccccc}M_1 & 0 & \frac{-g_1 v_d}{\sqrt{2}} & \frac{g_1 v_u}{\sqrt{2}} & 0\\
                       0 & M_2 & \frac{g_2 v_d}{\sqrt{2}} & \frac{-g_2 v_u}{\sqrt{2}} & 0\\
                       \frac{-g_1 v_d}{\sqrt{2}} & \frac{g_2 v_d}{\sqrt{2}} & 0 & -\mu_{eff} & -\lambda v_u\\
                       \frac{g_1 v_u}{\sqrt{2}} & \frac{-g_2 v_u}{\sqrt{2}} & -\mu_{eff} & 0 & -\lambda v_d\\ 
                       0 & 0 & -\lambda v_u & -\lambda v_d & 2\kappa s\end{array}\right)
\end{eqnarray}
R parity conservation in the scale invariant version of NMSSM puts LSP 
as a natural candidate of dark matter. Neutralino, as a linear combination of neutral components of
superpartners, see eq.(\ref{mix}), plays the role of LSP in most of the cases.
 \begin{equation}
  \arrowvert \tilde{\chi}_{1}^0\rangle = N_{11}\arrowvert \tilde{B} \rangle + N_{12}\arrowvert \tilde{W^0} \rangle 
   + N_{13}\arrowvert \tilde{H^0_d} \rangle + N_{14}\arrowvert \tilde{H^0_u} \rangle + N_{15}\arrowvert \tilde{S} \rangle
\label{mix}
 \end{equation}
In the neutralino, bino, wino, higgsinos and singlino-fractions are $N_{11}^2, N_{12}^2, N_{13}^2 + N_{14}^2$ 
and $N_{15}^2$ respectively.
\renewcommand{\arraystretch}{0.9} 
\begin{table}[h]
\centering
\begin{tabular}{|>{\centering\arraybackslash}m{5.5 cm}|>{\centering\arraybackslash}m{5.5 cm}|}\hline
Parameters & BM Point\\ \hline 
$N_{11}, N_{15} $ & 0.989, -0.039 \\ 
$N_{12}, N_{13}, N_{14}$ & 0.055, -0.123, 0.035 \\ 
$\sigma^p_{SI} \times 10^{-10}$ pb & 5.8 \\ 
$\sigma^p_{SD} \times 10^{-6}$ pb & 6.2 \\ 
$\sigma^n_{SI} \times 10^{-10}$ pb & 5.9 \\ 
$\sigma^n_{SD} \times 10^{-6}$ pb & 5.4 \\ 
\hline
\end{tabular}
\caption{The fractions of bino, wino, higgsinos and singlino in neutralino correspond to bench mark point of 
Table-(\ref{tab1}). 
The values of spin-independent and spin dependent cross-sections correspond to proton and neutron also shown, which is lying
outside the current limit of XENON100  \cite{Aprile2012} and can be tested in XENON1T \cite{Aprile2012a}}
\label{tab3}
\end{table}  
Using micrOMEGAs 3.1 code \cite{Belanger2013a} we find a bench mark point with parameters as shown in 
Table-(\ref{tab1}). This set of parameters give a SUSY spectrum (Table-\ref{tab1}), 
containing a 130 GeV bino dark matter with gauginos fractions shown in Table-(\ref{tab3}). 
It also contains a $\sim$ 125 GeV CP even Higgs as the lightest 
Higgs along with a light pseudo scalar Higgs of mass $\sim 68 ~\rm GeV$. Due to its singlet nature, this pseudo scalar Higgs 
can escape the lower bound of LEP \cite{Schael2006}.
The total decay width of Higgs and its branching ratios have also calculated using micrOMEGAs package. 
We mentioned the decay channels and branching ratios of lightest CP even Higgs in 
Table-(\ref{tab5}) and compared them with their Standard Model counterparts.
\renewcommand{\arraystretch}{0.8} 
\begin{table}[h]
\centering
 \begin{tabular}{|>{\centering\arraybackslash}m{3.5 cm}|>{\centering\arraybackslash}m{3.5 cm}
|>{\centering\arraybackslash}m{3.5 cm}|}\hline
Decay Processes & Branching Ratios (NMSSM) & $R = \frac{BR(H_i\rightarrow XX)}{BR(H_{SM}\rightarrow XX)}$\\[0.2 cm] \hline 
$H\rightarrow b\bar b$ & $6.57 \times 10^{-1}$  & 1.10 \\ [0.2 cm]
$H\rightarrow WW$ & $1.77 \times 10^{-1}$ & $\approx 0.90$ \\ [0.2 cm]
$H\rightarrow ZZ$& $2.072 \times 10^{-2}$ & $\approx 0.99$ \\ [0.2 cm]
$H\rightarrow \gamma\gamma$& $1.61 \times 10^{-3}$ & $\approx 0.69$ \\ [0.2 cm]
\hline
$\Gamma_{total}$& $4.479 \times 10^{-3}$ & $\frac{\Gamma^{i}_{total}}{\Gamma^{SM}_{total}} = 1.13$ \\ [0.2 cm]
\hline
\end{tabular}
\caption{Branching ratios in different channels and total decay width for NMSSM parameter space. $R$ gives the comparison of
these Branching ratios to their SM values.}
\label{tab5}
\end{table}     
The micrOMEGAs 3.1 code \cite{Belanger2013a} was also used to calculate the relic density of our bino dark matter and as it can
be seen from Table-(\ref{tab1}) that the value of relic density agrees with the WMAP-PLANCK \cite{Ade2013,Hinshaw2012} 
results i.e $\Omega_{\chi}h^2 = 0.1199 \pm 0.0027$.
 \begin{table}[h]
  \begin{center}
    \begin{tabular}{|>{\centering\arraybackslash}m{4 cm} | >{\centering\arraybackslash}m{2 cm} |}
    \hline
    Parameters at EW scale &   \\ \hline
    $\rm tan\beta$ & 1.65  \\ [0.5 cm] 
    \hline
    $\mu_{eff}$    & -208 \\ [0.5 cm]
    \hline
    $\lambda$      & -0.62 \\ [0.5cm]
    \hline
    $\kappa$       & -0.16  \\ [0.5 cm]
    \hline
    $A_{\lambda}$ [GeV] & -380  \\ [0.5 cm]
    \hline
    $A_{\kappa}$  [GeV] & -40    \\ [0.5 cm]
    \hline
    $M_{\tilde L_2}$ [GeV] & 152 \\ [0.5 cm]
    \hline
    $M_{\tilde \mu_R}$ [GeV] & 152  \\ [0.5 cm]
    \hline
    $M_{\tilde L_3}$ [GeV] & 180 \\ [0.5 cm]
    \hline
    $M_{\tilde \tau_R}$ [GeV] & 180  \\ [0.5 cm]
    \hline
    $M_1$     [GeV] & 129   \\ [0.5 cm]
    \hline
    $M_2$     [GeV] & 260    \\ [0.5 cm]
    \hline
    $M_3$     [GeV] & 1300     \\ [0.5 cm] 
    \hline
    \end{tabular}
%    \caption{Value of the parameters specified at EW scale}
%    \label{tab1}
% \end{center}
% \end{table}
    \quad
%     \renewcommand{\arraystretch}{1.1}
%     \begin{table}[h]
%     \begin{center}
    \begin{tabular}{|>{\centering\arraybackslash}m{4 cm} | >{\centering\arraybackslash}m{2 cm} |}
    \hline
    Mass Spectrum  & \\ \hline
    \multicolumn{2}{|l|}{SM-like Higgs Boson } \\[0.22 cm]
    \hline
    $M_{H_1}$ [GeV] &  125.32                     \\ [0.22 cm]
    \hline
    \multicolumn{2}{|l|}{Remaining Higgs spectrum} \\ [0.22 cm]
    \hline
    $M_{H_2}$ [GeV] &  150.68                         \\ [0.22 cm]
    \hline
    $M_{H_3}$ [GeV] &  453.09                          \\ [0.22 cm]
    \hline
    $M_{A_1}$ [GeV] &  67.47                             \\[0.22 cm]
    \hline
    $M_{A_2}$ [GeV] &  453.52                              \\ [0.22 cm]
    \hline
    $M_{H^{\pm}}$  [GeV] & 444.21                           \\ [0.22 cm]
    \hline
    \multicolumn{2}{|l|}{Sparticle masses and stop mixing } \\ [0.22 cm]
    \hline
    $m_{\tilde {g}}$   [GeV] & 1352.6        \\ [0.22 cm]
    \hline
    $m_{{\tilde \chi}_1^{\pm}}$ [GeV] & 216.8  \\ [0.22 cm]
    \hline
    $m_{{\tilde \chi}_2^{\pm}}$ [GeV] & 284.2    \\ [0.22 cm]
    \hline
    $m_{{\tilde \chi}_1^0}$     [GeV] & 130.1      \\ [0.22 cm]
    \hline
    $m_{{\tilde \chi}_2^0}$     [GeV] & 134.6       \\ [0.22 cm]
    \hline
    $m_{{\tilde \chi}_3^0}$     [GeV] & 233.5        \\ [0.22 cm]
    \hline
     $m_{{\tilde \chi}_4^0}$    [GeV] & 235.6         \\ [0.22 cm]
    \hline
     $m_{{\tilde \chi}_5^0}$    [GeV] & 285.7          \\ [0.22 cm]
    \hline
   \multicolumn{2}{|l|}{Relic Density} \\ 
   \hline
   $\Omega h^2 $ &  0.107   \\ [0.01 cm]
    \hline
    \end{tabular}
   \caption{The values of the parameters specified at electroweak scale and the output spectrum of SUSY particles.} 
   \label{tab1}   
   \end{center}
    \end{table} 
%%%%%%%%%%%%%%%%%%%%%%%%%%%%%%%%%%%%%%%%%%%%%%%%%%%%%%%%%%%%%%%%%%%%%%%%%%%%%%%%%%%%%%%%%%%%%%%%%%%%%%%%%%%%%%%%%%%%%%%%%%%%%%
\section{Gamma Ray Signal and Internal Bremsstrahlung}
% Neutralino, as a specific mixture of bino, wino, higgsinos
% and singlino decides the strength of $\gamma$-ray signal. 
% Approximate values of neutralino annihilation 
% cross-sections into $\gamma \gamma, \gamma Z$ have shown in Table-(\ref{tab4}) \cite{Shakya2012}.
% \begin{table}[h]
% \centering
%  \begin{tabular}{|>{\centering\arraybackslash}m{2.5 cm} | >{\centering\arraybackslash}m{2.5 cm}|
% >{\centering\arraybackslash}m{2.5 cm}|>{\centering\arraybackslash}m{2.6 cm}|>{\centering\arraybackslash}m{2 cm}|}
%              \hline
% Model & $\sigma_{\gamma\gamma}v (cm^3 s^{-1})$ &  $\sigma_{\gamma Z}v (cm^3 s^{-1}) $ & 
%       $\sigma_{total}v (cm^3 s^{-1}) $ & $R^{th}$\\[0.2 cm] \hline 
% Bino  & $\sim 10^{-30}$ & $10^{-31}$ & $10^{-27}$ & $\sim 1000$ \\ [0.2 cm]
% Wino  & $2.5 \times 10^{-27}$ & $1.4 \times 10^{-26}$  &  $4 \times 10^{-24}$ & 210 \\ [0.2 cm]
% higgsino & $1.1 \times 10^{-28}$ &  $3.7 \times 10^{-28}$ & $ 4.2 \times 10^{-25}$ & 710 \\ [0.2 cm]
% \hline
% \end{tabular}
% \caption{Limits on annihilation cross section and continuum ratio $R^{th}$for pure bino, wino and higgsino dark matter}
% \label{tab4}
% \end{table}
Recently, 130 GeV gamma ray signal coming from the vicinity of Galactic Center, has been observed in Fermi-LAT 
experiment \cite{Ackermann2012b} and it was found that a dark matter with mass
$129 \pm 2.4$ GeV \cite{Weniger2012} and cross-section 
$\langle\sigma v\rangle_{\gamma \gamma} = (1.27 \pm 0.32)\times 10^{-27} \rm cm^{3} sec^{-1}$ 
\cite{Weniger2012,Tempel2012a,Hektor2013a,Su2012,Su2012a,Hooper2012,Hektor2012,Mirabal2013,Zechlin2012} 
fits the signal very well.
% In Table-(\ref{tab4}), 
The wino or higgsino dark matter can satisfy these limits, but as suggested by  \cite{Cohen2012}, they will be ruled out 
by continuum constraints. A bino dark matter with
internal bremsstrahlung can explain the observed feature of Fermi-LAT experiment \cite{Ackermann2012b} and at the same
time avoids continuum constraints . 
% The enhanced annihilation cross-section can be obtained through internal bremsstrahlung, when the mass of neutralino 
% becomes degenerate with sleptons mass; assuming mass less leptons in final state. 
This idea will be discussed now in detail.\\
\begin{figure}[h]
\centering
\includegraphics[width=14cm,height=3cm]{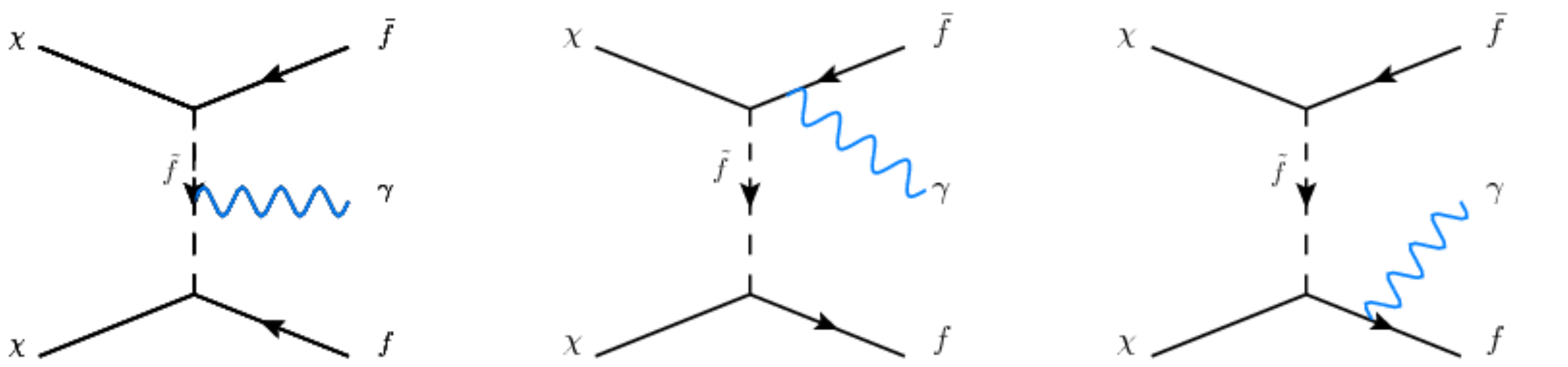}
\caption{Dark Matter annihilation into two fermion, first figure corresponds to virtual internal bremsstrahlung (VIB) and 
the other two correspond to Final state radiation (FSR), jointly known as internal bremsstrahlung}
\label{fig1}
\end{figure}
We take the scenario of Fig.(\ref{fig1}), where dark matter annihilates into two fermions. In this 
case, s-wave annihilation cross section will be helicity suppressed and p-wave contribution will be velocity suppressed 
\cite{Bringmann2012}. The presence of photon in the final state can 
lift the helicity suppression of $(m_f/m_{\chi})^2$ in the s wave annihilation cross-section, but suppress it 
by $\alpha_{em}$ \cite{Bergstrom1989,Flores1989}. 
In Fig.(\ref{fig1}), 
three possibilities have been 
shown, the first one correspond to virtual internal bremsstrahlung (VIB) and the last two correspond to final state radiation
(FSR), jointly known as internal bremsstrahlung (IB). 
The enhanced annihilation cross-section can be obtained through internal bremsstrahlung when the mass of neutralino 
is nearly degenerate with slepton mass and massless leptons are present in final state. 
The annihilation cross-section for the process
 $\chi \chi \rightarrow \bar f f \gamma$, where $\gamma$ is a virtual photon is given by
\cite{Bergstrom2008a,Bringmann2012,Bell2011},
 \begin{align}\label{cs0}
  \langle\sigma v\rangle_{v\rightarrow 0} &= \frac{\alpha_{em} |\tilde g_R|^4}{64 \pi^2 m^2_\chi}
  \Biggr\{\frac{3+4 \mu_R}{1+ \mu_R} + \frac{4 \mu^2_R - 3\mu_R -1}{2\mu_R} \rm log\frac{\mu_R-1}{\mu_R+1}\\\nonumber 
  &+ (1+\mu_R)\left[\frac{\pi^2}{6}-\left(\rm log\frac{\mu_R +1}{2\mu_R}\right)^2 - 2 \rm Li_2\left(\frac{\mu_R+1}{2\mu_R}
  \right)\right]\Bigg\} + (R\leftrightarrow L)
 \end{align}
where ${\rm Li_2}(z)= \Sigma^{\infty}_{k=1} z^k/k^2$, $\mu_{R,L}\equiv m^2_{\tilde f_{R,L}}/ m^2_{\chi}$ 
and $\tilde g_R (\tilde g_L)$ is the coupling between neutralino, leptons and sleptons. This expression is valid for the 
massless 
fermions in the final state, within MSSM scenario. We used micrOMEGAs$3.1$ \cite{Belanger2013a} code for our dark matter 
calculations and found that bino is the possible dark matter for our bench mark point, see Table-(\ref{tab3}). Since bino is 
the dark matter 
candidate, we can use the same annihilation cross-section of eq.(\ref{cs0}), for our calculation. The Yukawa couplings in 
MSSM scenario, can be defined as \cite{Haber1985}.
\begin{equation}
 {\tilde g_L} = -\frac{2Q_f \mp 1}{\sqrt{2}} g_2~{\rm tan\theta_W}~N_{11} \mp \frac{g_2}{\sqrt{2}}~N_{12}
 \label{coup1}
 \end{equation}
 \begin{equation}
 {\tilde g_R} = \sqrt{2} Q_f~g_2~{\rm tan\theta_W}~N_{11}
 \label{coup2}
\end{equation}
where $g_2$ is the $SU(2)$ coupling and $\theta_W$ is the Weinberg angle. $N_{11}$ and $N_{12}$ has defined in 
eq.(\ref{mix}) and their values have given in Table.(\ref{tab3}). 
In eq.(\ref{coup1}), $\mp$ signs correspond to isospin, $T_3 = \pm \frac{1}{2}$ respectively. In our bench 
mark scenario, the values of the couplings as given by eq.$(\ref{coup1},\ref{coup2})$ are,
\begin{equation}
 \tilde g_L \approx 0.27,~\tilde g_R \approx -0.50
\end{equation}
After using these values with corresponding $N_{11}$ and $N_{12}$ from Table.(\ref{tab3}), annihilation 
cross-section for electron-positron final state comes,
\begin{equation}
\langle\sigma v\rangle \approx 2.4\times 10^{-29} \rm cm^3/sec
\end{equation}
which is less than the required cross-section ($O\sim 10^{-27}$ \cite{Weniger2012}), for fitting the signal of 130 GeV dark
matter and we need a boost factor of $\sim 590$ to explain Fermi-LAT data  \cite{Ackermann2012b}.
%%%%%%%%%%%%%%%%%%%%%%%%%%%%%%%%%%%%%%%%%%%%%%%%%%%%%%%%%%%%%%%%%%%%%%%%%%%%%%%%%%%%%%%%%%%%%%%%%%%%%%%%%%%%%%%%%%%%%%%%%%%%%
\begin{figure}[h]
\centering
\includegraphics[width=13cm,height=8cm]{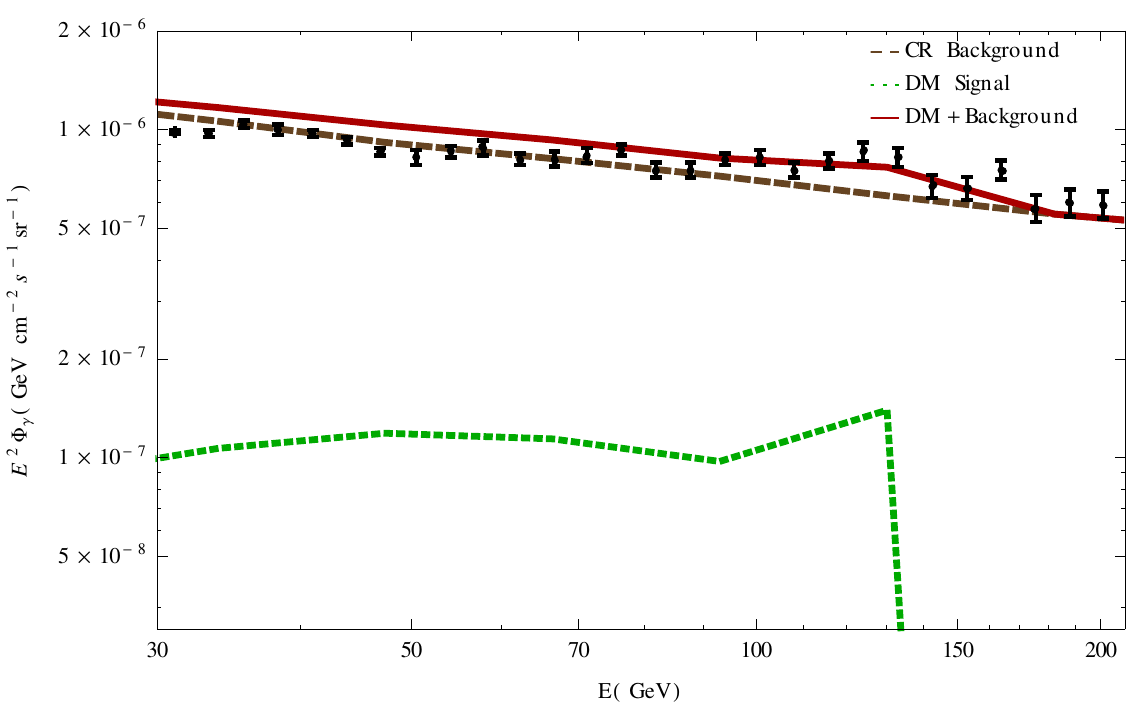}
\caption{Diffuse $\gamma$-ray flux for 130 GeV, bino dark matter in the bench mark scenario of Table-(\ref{tab1}). FERMI-LAT 
data \cite{Ackermann2012b} has been shown for comparison.}
\label{fig4}
\end{figure} 
\section{Indirect Detection}
Indirect detection of dark matter relies on the observations of its annihilation products. There are various experiments like
IceCube \cite{Abbasi2012}, PAMELA \cite{Adriani2009} and Fermi-LAT \cite{Ackermann2012b}, which are looking for the different
form of annihilation products of dark matter. Recently, AMS-02 experiment \cite{Aguilar2013} observed an excess in 
positron flux, which may be the indirect signal of dark matter, but to satisfy AMS-02 data, one needs a $\sim$ TeV 
range dark matter, which is not the case for our bench mark point. Hence in this work, we do not attempt to fit the 
positron excess observed in AMS-02 experiment \cite{Aguilar2013}. 
In section-III, we discussed the effect of internal 
bremsstrahlung on the dark matter annihilation cross-section, since dark matter is a majorana particle, its cross section 
for the process of eq.(\ref{ann1}), will be suppressed by helicity conservation.
\begin{equation}
 \chi \chi \rightarrow f \bar f \gamma
 \label{ann1}
\end{equation}

\begin{figure}[h]
\centering
\includegraphics[width=13cm,height=9cm]{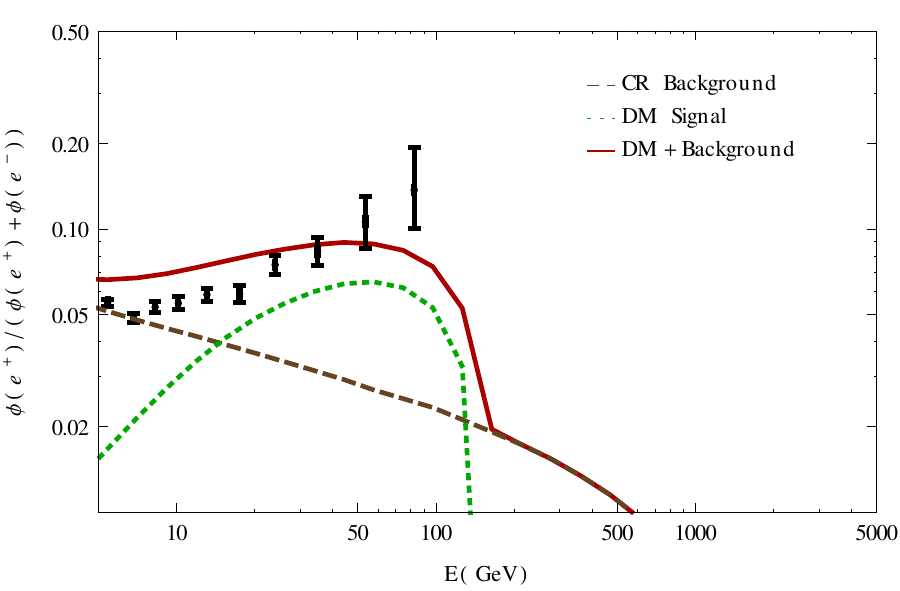}
\caption{Ratio of the positron flux $\phi(e^{+})$ to the total flux $(\phi(e^{+}) + \phi(e^{-}))$ vs Energy for 130 GeV, bino
dark matter in the bench mark scenario of Table.(\ref{tab1}). Background and dark matter signals with PAMELA data 
\cite{Adriani2009} have been shown for comparison.}
\label{fig5}
\end{figure}

\begin{figure}[h]
\centering
\includegraphics[width=13cm,height=9cm]{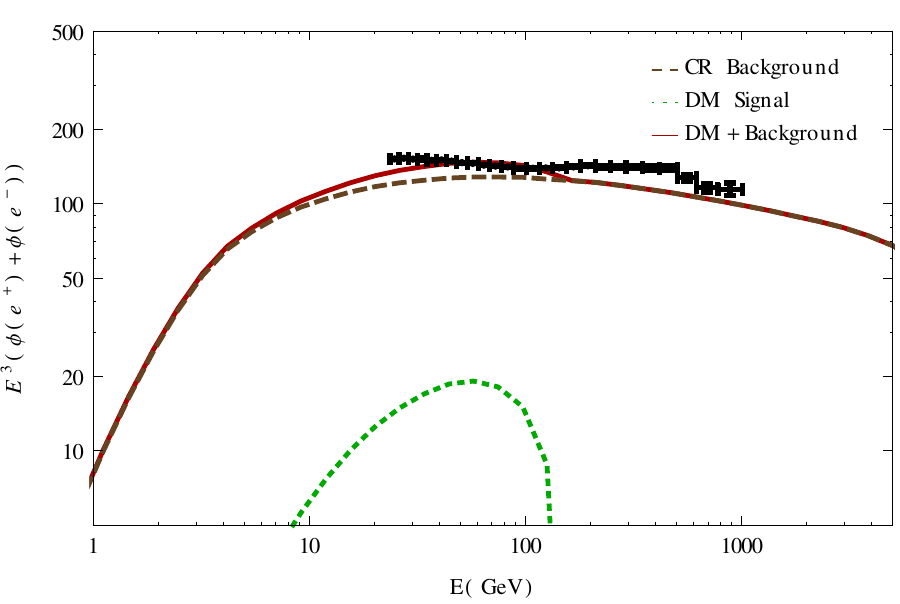}
\caption{The $(\phi(e^+) + \phi(e^-))$ flux for 130 GeV bino dark matter, Fermi-LAT data \cite{Abdo2009,Ackermann2010} has been shown
for comparison. Green dotted line denotes the dark matter signal and Brown dashed line denotes the background}
\label{fig6}
\end{figure}

\begin{figure}[h]
\centering
\includegraphics[width=13cm,height=9cm]{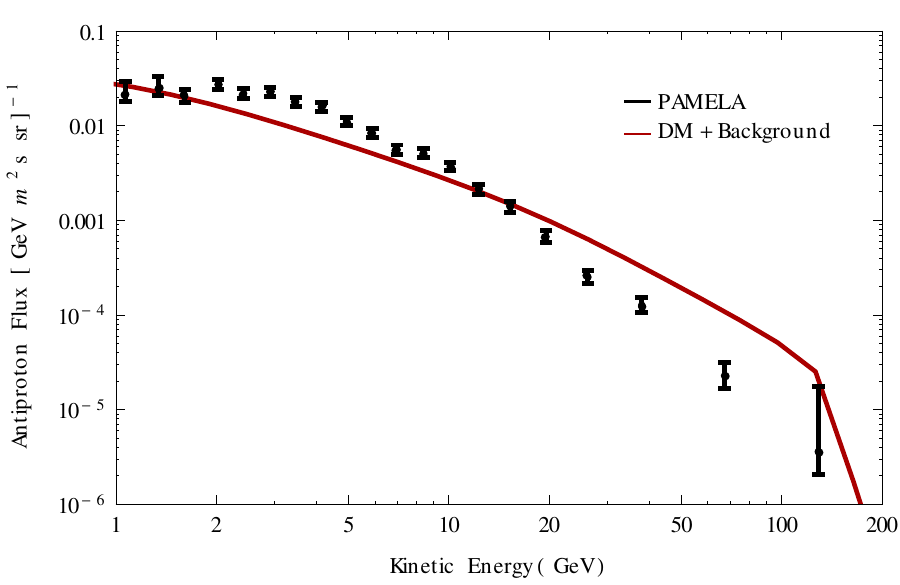}
\caption{The $(e^+ + e^-)$ flux for 130 GeV bino dark matter, Fermi-LAT data \cite{Abdo2009,Ackermann2010} has been shown
for comparesion. Green dotted line denotes the dark matter signal and Brown dashed line denotes the background}
\label{fig7}
\end{figure}

If there is a photon present in the final state, as shown in Fig.(\ref{fig1}), helicity suppression can be avoided, 
leaving behind 
$\alpha_{em}$ suppression, which gives a much larger cross-section. 
We got photon flux from micrOMEGAs 3.1 code \cite{Belanger2013a} for our bench mark scenario and used this flux with  
boosted ($\sim 590$) annihilation cross section in GALPROP code \cite{Strong2007, Moskalenko1998}.
In Fig.(\ref{fig4}), GALPROP output for diffuse gamma ray flux using isothermal
density profile \cite{Bahcall1980} has been plotted as a function of energy. It can be seen from Fig.(\ref{fig4}) that 
there is a clear bump at 130 GeV, which satisfy the observed feature of Fermi-LAT data \cite{Ackermann2012b}. 
Since, a boost factor of $\sim 590$ is required to explain Fermi-LAT data \cite{Ackermann2012b}, we explored the effects of 
this boost factor on the 
positron, electron and antiproton flux, that we will discuss now in details.\\
PAMELA satellite \cite{Adriani2009}, which detects antimatter, has observed an excess in the differential ratio 
$\frac{\phi(e^{+})}{\phi(e^{+} + e^{-})}$, which seems to indicate the presence of dark matter annihilation products.
As in the case of gamma, we use the electron, positron and antiproton fluxes obtained from micrOMEGAs, in the GALPROP
\cite{Strong2007, Moskalenko1998} with a boosted annihilation cross-section as before. 
The output spectrum of positron obtained from GALPROP, relative to electron for 130 GeV bino dark matter has been plotted in 
Fig.(\ref{fig5}). We have also plotted PAMELA data 
for comparison and found that boost factor of $\sim 590$ agrees with the positron excess observed in 
PAMELA data \cite{Adriani2009}. 
PAMELA data can be explained with a large boost factor that has been discussed in detail \cite{Bergstrom2008a,Mohanty2012}, 
where authors suggest, the presence of massive black hole as a reason for the large boost factor. 
In \cite{Bergstrom2008a}, flux ratio has been plotted for different mass 
range of dark matter and suggested that neutralino with mass greater than 100 GeV can fit the signal well, but the shape of
the signal gets flatter with increasing mass of dark matter. This is the case for our bench mark point, which satisfying Fermi
-LAT data under the constraint of PAMELA experiment. 
The low energy region of the plot in Fig.(\ref{fig5}), seems to be in disagreement with the observed data, but due to 
solar modulation
effect, the constraint of PAMELA experiment in this region is relaxed.
\\
In addition we also do not see any excess in the $(\phi(e^+) + \phi(e^-))$ flux observed by 
Fermi-LAT \cite{Abdo2009,Ackermann2010} experiment, as shown in Fig.(\ref{fig6}). And similarly no excess is observed in 
the case of antiproton flux observed by PAMELA \cite{Adriani2009b,Adriani2010} as shown in Fig.(\ref{fig7}).  
%%%%%%%%%%%%%%%%%%%%%%%%%%%%%%%%%%%%%%%%%%%%%%%%%%%%%%%%%%%%%%%%%%%%%%%%%%%%%%%%%%%%%%%%%%%%%%%%%%%%%%%%%%%%%%%%%%%%%%%%%%%%%
\section{Direct Detection}
\begin{figure}[h]
\centering
\includegraphics[width=13cm,height=4cm]{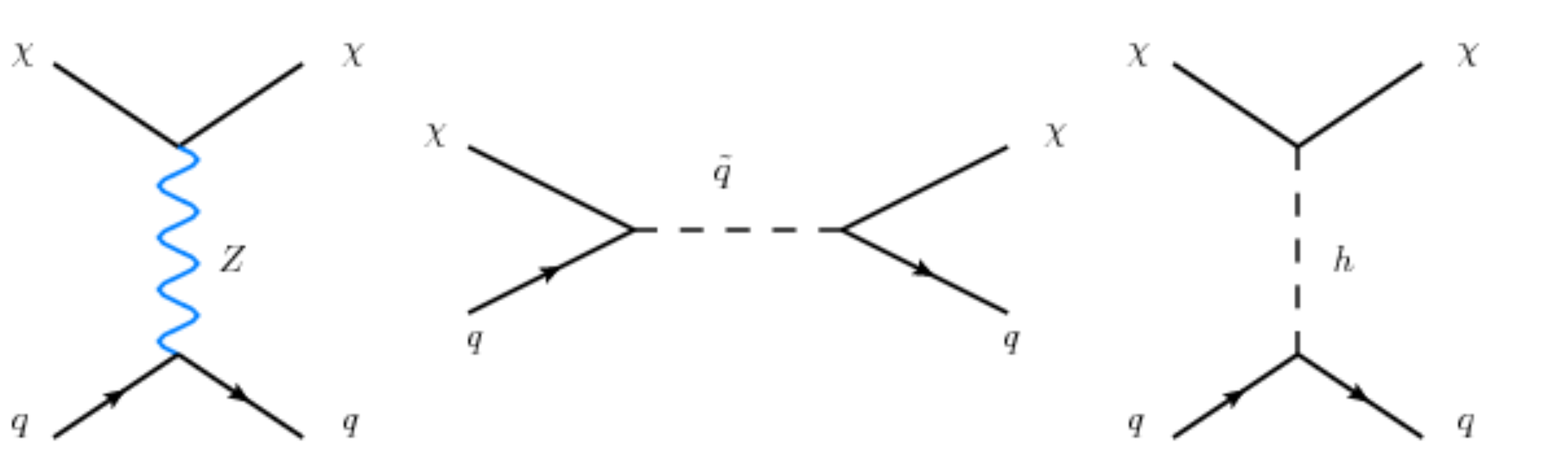}
\caption{Feynman diagrams for spin dependent and independent elastic scattering of neutralino from quarks.}
\label{fig2}
\end{figure}

In the direct detection of Weakly Interacting Massive Particle (WIMP), elastic scattering cross-section of WIMP from a 
heavy nuclei like Xenon or Germanium, plays
an important role. There are possibilities for different types of interactions (Fig.\ref{fig2}) between WIMP and matter nuclei, 
but two of them play the major role. Spin-spin interaction is one of the important interaction, 
where WIMP couples to the spin of the nucleus. 
The other important coherent interaction is scalar interaction, in which WIMP couples to the mass of the nucleus.
The low energy effective Lagrangian for these two interactions is \cite{Jungman1996a},
\begin{equation}
 {\cal L}_{eff} = c_q {\bar \chi} \chi {\bar q} q + d_q {\bar \chi} \gamma^{\mu} \gamma_5 \chi \bar q \gamma_{\mu} \gamma_5 q
 \label{eff}
\end{equation}
where $c_q$ and $d_q$ are the couplings corresponding to spin independent and dependent interactions respectively. As pointed
out in \cite{Chalons2013}, these couplings are model dependent and in NMSSM, $c_q$ will be proportional to the
coupling of neutralino-neutralino to Higgs. Since in our case squarks are heavy ($\sim$ TeV), so they decouple easily 
and don't play any significant role in the scattering processes.\\
\begin{figure}[h]
\centering
\includegraphics[width=13cm,height=8cm]{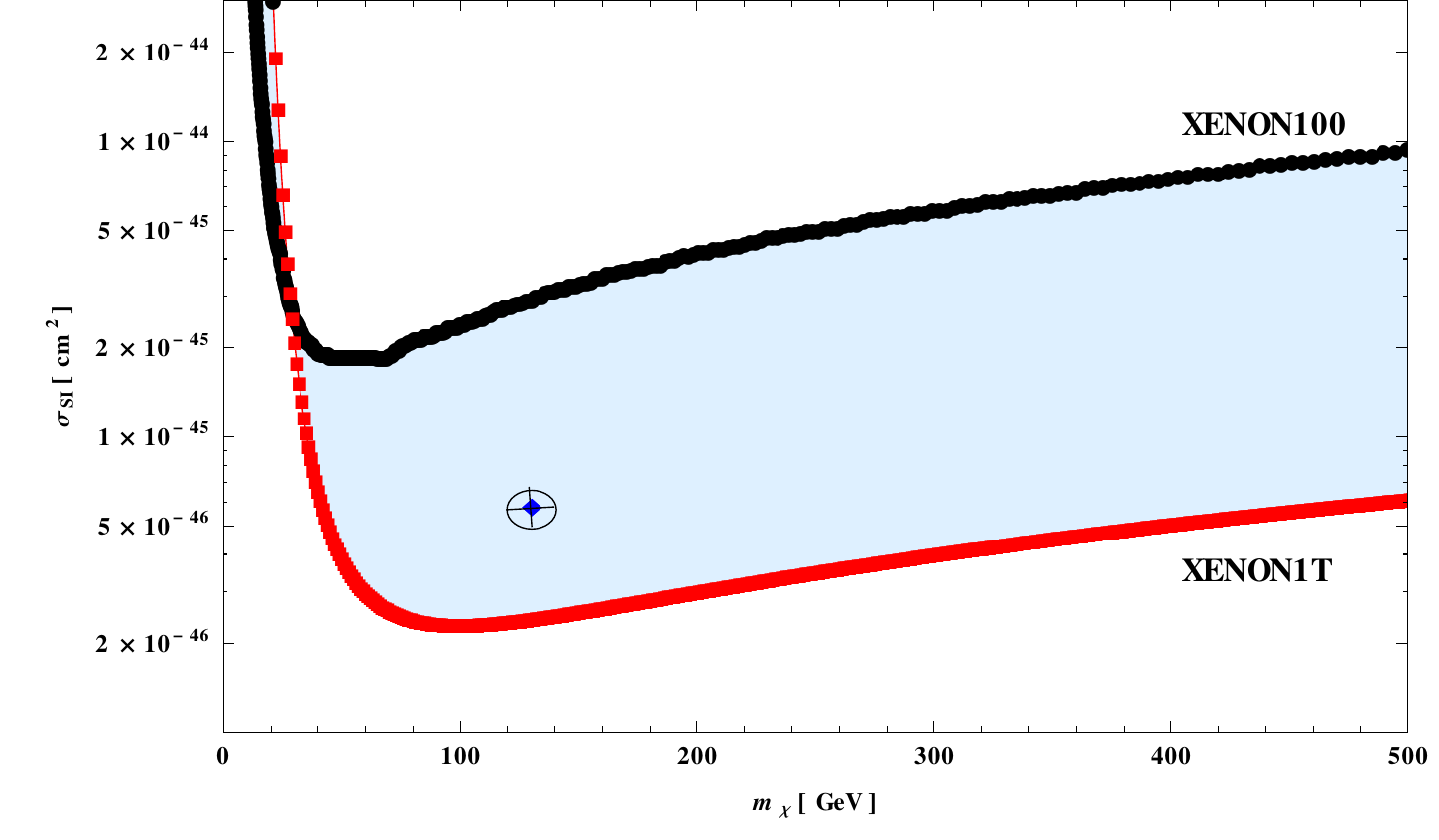}
\caption{Plot of spin independent cross-section versus neutralino mass. XENON100 \cite{Aprile2012} and 
XENON1T \cite{Aprile2012a} data has shown for comparison.}
\label{fig3}
\end{figure}
We use micrOMEGAs 3.1 code \cite{Belanger2013a}, for the calculation of spin dependent and independent cross-section for
our bench mark scenario. We find scalar interaction cross-section, 
$\sigma^p_{\rm SI} = 5.8\times 10^{-10}$ pb for $\sim 130$ GeV neutralino. This cross-section goes beyond the current 
limit of XENON100 experiment, which is $\sigma^p_{\rm SI} \sim 3\times 10^{-9}$ pb for 
130 GeV neutralino \cite{Aprile2012}. In Fig.(\ref{fig3}), We plot $\sigma^p_{\rm SI}$ as a function of $m_{\chi}$ along 
with the exclusion 
limit from XENON100 \cite{Aprile2012} and XENON1T \cite{Aprile2012a} experiments. 
It is clear that our point evades 
the current constraint of XENON100 and can be tested in XENON1T experiment.\\
Spin dependent interaction is mediated by $Z$ boson and for scale invariant NMSSM, it can be given as \cite{Chalons2013},
\begin{equation}
 \sigma^p_{\rm SD} \approx 4.0 \times 10^{-4} {\rm pb} \left(\frac{|N_{13}|^2 - |N_{14}|^2}{0.1}\right)^2
\label{sd}
\end{equation}
In our case, bino is the LSP and $N_{13}$, $N_{14}$ components are very small, so it is easy to evade present bounds on 
spin dependent cross-section coming from Super-Kamiokande and IceCube experiments \cite{Tanaka2011, Abbasi2012}, 
which is $\sim 2.7 \times 10^{-4}$pb for $\sim100$ GeV neutralino. Our result
of
$\sigma^p_{\rm SD} = 6.21 \times 10^{-6}$pb for $m_{\chi^0} \sim 130$ GeV is close to the value coming from 
analytical expression of eq.(\ref{sd}) and evades the present bounds. All the values of bino, wino, higgsino and
singlino fractions with spin dependent 
and independent cross-sections are shown in Table-(\ref{tab3}). This result can be further tested in future experiments 
\cite{Aprile2012a,Fiorucci2013}.
% %%%%%%%%%%%%%%%%%%%%%%%%%%%%%%%%%%%%%%%%%%%%%%%%%%%%%%%%%%%%%%%%%%%%%%%%%%%%%%%%%%%%%%%%%%%%%%%%%%%%%%%%%%%%%%%%%%%%%%%%%%%%%
% \section{LHC Perspective}
% In this section, we will discuss our bench mark scenario in the perspective of LHC. 
% It is clear from Table-(\ref{tab1}) that
% the masses of gauginos, squarks and sleptons with their trilinear couplings respect the current bounds of 
% ATLAS \cite{ATLAS2012} and CMS \cite{Chatrchyan2012a} experiments. We are using the recent version of micrOMEGAs 3.1 
% \cite{Belanger2013a}, in
% which an interface to the HiggsBounds, which is a publicly available code \cite{Bechtle2011}, has been provided. HiggsBounds 
% \cite{Bechtle2011}
% takes the present experimental (ATLAS \cite{ATLAS2012}, CMS\cite{Chatrchyan2012a}, LEP \cite{Schael2006}etc.) 
% constraints into account and accordingly include or exclude different 
% channels. The output of HiggsBounds signifies that our bench mark point is completely valid in the present scenario. 
% For our bench mark point, $\mu_{eff}$ is very close to the limit 200, which is required for tree level naturalness and the 
% vale of $\rm tan\beta$ is also small that gives the maximum contribution to the Higgs mass from tree-level.\\
% Mass spectrum of Table-(\ref{tab1}), has been obtained using micrOMEGAs \cite{Belanger2013a}, which calculates this with the help 
% of NMSSMTools package.
%%%%%%%%%%%%%%%%%%%%%%%%%%%%%%%%%%%%%%%%%%%%%%%%%%%%%%%%%%%%%%%%%%%%%%%%%%%%%%%%%%%%%%%%%%%%%%%%%%%%%%%%%%%%%%%%%%%%%%%%%%%%%%%
\section{Conclusion}
In this paper, we explored the internal bremsstrahlung in the singlet extension of MSSM and found that it can give
an explanation for recently observed feature in Fermi-LAT data. We discussed the internal bremsstrahlung in the NMSSM
and fit the Fermi-LAT data for isothermal profile of dark matter.
We have chosen light sleptons such as they lie near to the mass
of dark matter to get an enhanced cross-section. In the internal bremsstrahlung case, 
to satisfy Fermi-LAT data, we
need a boost factor of $\sim 590$. The effect of the boost factor on the flux of electrons, positrons and antiprotons 
have been 
discussed in details and we do not observe any excess in the spectrum of electrons, positrons and antiprotons, 
over the CR background, as shown in Fig.(\ref{fig5},\ref{fig6},\ref{fig7}). We also discussed the direct and indirect 
detection consequence of our dark matter scenario. 
It is clear from Fig.(\ref{fig3}) that for spin independent 
cross-section, our bench mark point evades the current bound of XENON100 easily and can be tested in future experiments. 
Spin dependent cross-section also evades the present experimental bound and can be verified in future experiments.\\
\bibliography{Tomar}
\bibliographystyle{apsrev4-1.bst}
\end{document}